\documentclass[aps,twocolumn,showpacs,preprintnumbers,prc,nofootinbib]{revtex4-2}
\usepackage{graphicx,color}
\usepackage{float}
\usepackage{amsmath,amssymb,amsfonts}
\usepackage{bm}
\usepackage[pdftex,colorlinks=true, linkcolor = blue, citecolor=blue,urlcolor=blue, bookmarksnumbered=true, bookmarksopen=true]{hyperref}
\usepackage[normalem]{ulem}

\newcommand{\ket}[1]{\lvert #1 \rangle}

\begin{document}

\title{Antisymmetrization of composite fermionic states for quantum simulations of nuclear reactions in first-quantization mapping}

  \author{Ionel Stetcu}
\affiliation{
  Theoretical Division, Los Alamos National Laboratory, Los Alamos, New Mexico 87545, USA}
  
\date{\today}
\preprint{LA-UR-25-32003}

\begin{abstract}
I present a first-quantization deterministic algorithm for antisymmetrizing a spatially separated target–projectile system containing $N_T$ and $N_p$ identical fermions, respectively. The method constructs a fully antisymmetric wavefunction from the product of two independently antisymmetrized many-body states, each of which may be a superposition of Slater determinants. The algorithm uses a Dicke-state ancilla register that coherently encodes all one-particle exchange channels between the two subsystems and, crucially, requires only \emph{single-particle} swaps to generate the full antisymmetric structure. After consolidating distinct control conditions using logical \textsc{or} operations, the optimized circuit uses \(N_TN_p\) single-particle exchanges, up to \(N_p\) of which can be performed in parallel. The correct fermionic phase is incorporated through the application of $Z$ gates on $N_T$ ancillae, after which the ancilla register is efficiently uncomputed using a compact sequence of controlled operations. This construction provides a nontrivial and scalable protocol for preparing fully antisymmetric states in reaction and scattering simulations, significantly expanding the range of systems that can be addressed with first-quantized quantum algorithms.
\end{abstract}
\maketitle

%\section{Introduction}

Progress in simulating nuclear dynamics on classical computers has historically been incremental, with controlled approximations achievable only in specific kinematic regimes or for limited few-body systems. Quantum computing offers a promising alternative, as it can, in principle, provide an exponential advantage for solving quantum many-body dynamics. This advantage arises from the ability of a quantum processor to emulate the unitary time evolution generated by a many-body Hamiltonian, thereby enabling direct simulations of nuclear dynamics with reduced reliance on approximations. In practice, however, even fault-tolerant quantum computers will introduce inherent errors, so the resulting evolution will only approximate the exact unitary behavior. Nevertheless, steady progress in quantum error correction \cite{Katabarwa2023EFTQC,Wang2025SAQEC,Zeng2025ESTQC,Senior2025NeuralDecoder,Aasen2025FTQC,quera:2025} and hardware fidelity \cite{Smith2025FluxoniumFidelity,Loeschnauer2025AllElectronicTI,Salmanogli2025CoDesign} provides growing confidence that such errors will remain controllable, opening the path to realistic quantum simulations of nuclear reactions.

Nuclear reaction cross sections exemplify the difficulty of describing and measuring quantum many-body dynamics in complex systems. As the incident energy varies, multiple competing reaction channels open, creating an intricate landscape that is challenging to disentangle theoretically. Experimentally, closely spaced states and overlapping thresholds hinder the isolation of specific processes, especially near channel openings where signals are weak and strongly mixed. In heavy nuclei, the coexistence of multiple decay modes and the emission of secondary particles further obscure channel identification, making precise and unambiguous cross section measurements difficult even in seemingly simple cases.

Theoretically, approaches ranging from R-matrix methods \cite{Descouvemont:2010} to Hauser–Feshbach models \cite{HF1952} rely on significant approximations and are most reliable when constrained by data; in data-sparse regimes, predictions can vary widely across implementations. Fully microscopic calculations based on realistic interactions remain limited to light nuclei
\cite{PhysRevLett.99.022502,PhysRevLett.101.092501,Leidemann2009,PhysRevC.79.044606,ATKINSON2025139189,PhysRevC.108.034001},
and often additional approximations are required. These limitations point to quantum computing as a promising avenue to overcome the prohibitive scaling of classical many-body methods and enable more direct \textit{ab initio} treatments of nuclear reaction dynamics. The approach recently proposed in Ref.~\cite{Rule:2025scatt} offers a pathway toward this objective, and the present work contributes one of the key implementation tools needed to realize it.

The quantum simulation of nuclear dynamics generally involves three major components: (1) state preparation, in which the quantum register is initialized to represent the physical system of interest; (2) time evolution, where the system evolves under a unitary operator derived from the underlying nuclear Hamiltonian; and (3) measurements, which are used to extract observables such as scattering amplitudes, transition probabilities, or cross sections. For scattering and reaction studies, where the correct asymptotic behavior of the wave function is essential, it is often more efficient to employ the first-quantization mapping, in which the number of required qubits scales linearly with the number of particles and only logarithmically with the number of available single-particle states. This makes first quantization particularly well suited for describing reactions in extended spatial volumes and for extracting asymptotic observables. 

Since the introduction of quantum computing techniques to nuclear physics~\cite{PhysRevLett.120.210501}, 
a growing body of work has explored their application to problems in nuclear structure~\cite{PhysRevC.104.024305,Stetcu2022var,Kiss:2022,perezobiol2023nuclear,Stetcu-2023proj,Li:2024} 
and to the computation of scattering observables~\cite{Rogerro2019,Sharma:2023bqu,Turro:2024ksf,Wang:2024scd,Weiss:2024mie,Spagnoli:2025}. 
Although nuclear structure calculations are comparatively more straightforward to adapt to quantum 
algorithms and can yield near-term progress, the simulation of nuclear dynamics and reactions requires 
further algorithmic development. Large-scale reaction simulations will rely on the 
integration of several algorithmic components, including state preparation, time evolution, 
antisymmetrization, and measurement protocols for extracting physical observables. Continued advances 
in these areas are expected to ultimately enable a paradigm shift in the simulation of nuclear dynamics, 
including direct calculation of reaction cross sections.

\begin{figure*}
    \centering
    \includegraphics[width=0.85\linewidth]{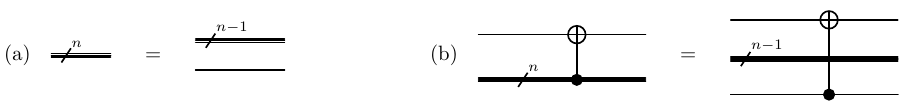}
    \caption{Encoding of particle states and controlled operations in the quantum-circuit representation. (a) An $n$-qubit particle state, where the last qubit labels the particle type, with $\ket{0}$ denoting the target and $\ket{1}$ denoting the projectile. (b) A \textsc{cnot} operation controlled on the particle state, implemented as a controlled operation acting on the last ancilla qubit associated with that particle state, used for uncomputing ancillae.}
    \label{fig:equations-cnot}
\end{figure*}

In the context of reaction dynamics, it will be necessary to prepare two initial many-body states, one describing the target and another describing the projectile, each antisymmetrized internally and subsequently antisymmetrized with respect to one another to ensure full fermionic symmetry. State preparation poses a particularly significant challenge for fermionic systems, as it requires the use of unitary transformations that can be challenging. Although a unified projection framework for constructing both subsystems is in principle feasible, practical differences in their structure, which determine the optimal evolution time and outcome probability of the projection, make a two-step strategy more efficient, in which the antisymmetrization between target and projectile is performed only after the projected states have been obtained. If the target and projectile states are prepared simultaneously, the overall success probability is the product of the individual preparation probabilities. In contrast, if the two subsystems can be prepared independently, a successfully prepared state can be retained while the preparation of the other subsystem is repeated until successful, avoiding this multiplicative reduction in the overall success probability. This, in turn, motivates the algorithmic developments presented in this work, which address the antisymmetrization of such composite quantum states on a quantum computer. An additional application of this approach is the antisymmetrization of protons and neutrons, when it is advantageous to treat isospin as an explicit quantum number within a unified fermionic framework.

Several algorithms for constructing antisymmetric fermionic states have been 
proposed in the literature~\cite{AbramsLloyd1997,Berry2018a,Babbush:2023sas,10.1063/5.0239980,Rule:2025asym}. 
These techniques can be used to prepare initial states for two nuclei, which are 
subsequently refined through quantum phase estimation or related projection 
methods  \cite{Kitaev1995AbelianStabilizer,10.1063/1.5027484,Dong2022,PhysRevResearch.4.033121,Stetcu-2023proj,QPE-taper}, possibly combined with variational techniques \cite{Grimsley2019ADAPTVQE,Tang2021QubitADAPTVQE,Grimsley2023ADAPTlandscape}, including symmetry aware ansatzes \cite{Gibbs2025,Mihalikova:2025statepreparationsymmetries}, to enhance the probability 
of preparing ground or low-lying excited states of the target and projectile. At this stage, however, the product of the two eigenstates is not antisymmetric under the exchange of identical fermions across the subsystems, which is where the current work becomes essential. I further note that when the two subsystems are initialized at large spatial  separations, as is typical in reaction simulations, the full antisymmetrization does not modify their internal structure.

Most existing antisymmetrization schemes rely on ordered single-particle states~\cite{AbramsLloyd1997,Berry2018a,Babbush:2023sas,10.1063/5.0239980}, and some are probabilistic, although the probability of success can be significant. In contrast, \citet{Rule:2025asym} introduced a recursive deterministic algorithm that builds the antisymmetric wavefunction of $N$ particles iteratively from the antisymmetric state of $N-1$ particles through controlled-swap operations and uncomputation steps that enforce the correct antisymmetric phase. A measurement-based variant reduces the circuit depth through conditional phase corrections based on mid-circuit measurements. The algorithm supports arbitrary orthogonal single-particle states, requires no ordered inputs, and remains deterministic, enabling scalable quantum simulations of fermionic systems and nuclear reactions. 

Here, I generalize this approach to construct a fully antisymmetric state from the 
product of two independently antisymmetrized subsystems, each of which may be a 
superposition of multiple Slater determinants. The only requirement is that the 
subsystems remain readily distinguishable by inspecting a small number of qubits, 
which allows the procedure to avoid applying full inverse single-particle 
preparation circuits, in contrast to Ref.~\cite{Rule:2025asym}. As in 
Ref.~\cite{Rule:2025asym}, the extension introduced in this work can be viewed as a 
unitary analogue of projection operators commonly employed in nuclear many-body 
theory; by preserving the full many-body wave function, this formulation is 
naturally suited for quantum simulation and, in particular, for applications to 
nuclear reaction dynamics. From a classical perspective, the present approach is 
in the same spirit as the antisymmetrization method introduced by 
\citet{Navratil:2000as} in the Jacobi-coordinate formulation of the no-core shell 
model \textit{ab initio} method.

%\section{quantum algorithm}

Let us begin by assuming that we have prepared two initial states: one for the target consisting of \( N_T \) identical fermions and one for the projectile consisting of \( N_p \) identical fermions, with $N_p\leq N_T$. These systems are initially placed far apart so that at least one qubit can be used to distinguish between single-particle states belonging to the target and the projectile.

On the lattice, we can picture the target’s center of mass (CoM) localized on the left side and the projectile’s CoM on the right, such that the corresponding single-particle wavefunctions are spatially well separated. Furthermore, one can assume that one of the qubits describing the position on the lattice is in state $|0\rangle$ if located on the left hand side of the lattice and in state $|1\rangle$ if located on the right hand side. This will allow us to quickly identify the nucleons of the target or projectile, since identifying the single-particle states would be too complicated, especially when we want to antisymmetrize complicated many-body wave functions, and would increase the complexity of the algorithm. In quantum circuit language, this means that for a particle state represented on $n$ qubits, the last one will be assigned to control the target or the projectile, and in this work I will arbitrarily choose $|0\rangle$ if the particle state pertains to the target, and $|1\rangle$ if the particle state pertains to the projectile, as shown in Fig. \ref{fig:equations-cnot}(a).
For the purpose of uncomputing the ancillae, a \textsc{not} operation controlled on the particle state consists of a controlled operation on the last ancilla of that particle state, illustrated in Fig. \ref{fig:equations-cnot}(b),
and similarly if the operation is controlled on $\ket{0}$. Swaps between particle states are composed of swaps between individual qubits with the same rank.

\begin{figure*}
    \centering
    \includegraphics[width=0.75\linewidth]{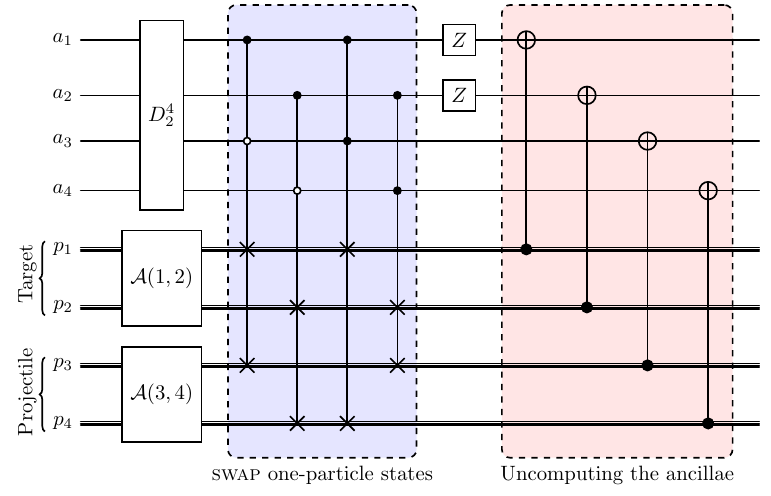}
    \caption{ Quantum circuit for antisymmetrizing the two-particle target and projectile states. Starting from independently antisymmetrized two-fermion states for the target and projectile and four ancilla qubits prepared in the Dicke state $D_2^4$, the blue-filled region performs all single-particle exchanges between the subsystems in parallel using two additional ancillae not shown here, with appropriate $Z$ gates providing the correct antisymmetrization phase. The swaps entangle projectile states with ancillae in state $\ket{1}$ and target states with ancillae in state $\ket{0}$, generating all required permutations. The uncomputation stage, shown in the pink-filled region, disentangles the ancilla register using $N_T+N_p$ controlled-\textsc{not} operations. %Redundant swaps are present but do not affect the final antisymmetrized state.
 }
    \label{fig:antisymm2}
\end{figure*}

%To fully antisymmetrize the combined state, one must consider all possible particle exchanges between the target and projectile, which amount to
%\begin{equation}
%    N_\mathrm{perm} = \frac{(N_T + N_p)!}{N_T! \, N_p!}
%    = 
%    \begin{pmatrix}
%        N_T + N_p \\
%        N_p
%    \end{pmatrix}
%    \label{eq:Nperm1}
%\end{equation}
%distinct terms.

For the case $N_p = 1$, corresponding, for example, to a neutron-, deuteron- or triton-induced reaction (in the case of triton, the neutrons are a bit more complicated to antisymmetrize, but the algorithm will be covered below), the algorithm of Ref.~\cite{Rule:2025asym} generalizes straightforwardly, since only a single projectile particle must be antisymmetrized with the $N_T$ target fermions. Although uncomputing the ancillae formally requires identifying proper components entangled with the ancilla register, the inverse state-preparation circuit for the target need not be applied. Instead, the procedure can be simplified by examining only the qubit that specifies the projectile’s position on the right side of the simulation box. Thus, the phase can be applied using a $Z$ gate on the special qubit for the particle where correction is necessary, that is, if a projectile state is moved into a target state.

For two or more particles in the target and projectile, the number of permutations that need to be implemented grows rapidly with the number of identical fermions in the target and projectile. Hence, one needs a different strategy from Ref.~\cite{Rule:2025asym}, even though one would like to preserve the simplicity of only implementing one-particle exchanges at a time. This can be achieved by using $N_T + N_p$ ancilla qubits and preparing them in the Dicke state $D^{N_T+N_p}_{\,N_p}$~\cite{Dicke1954Coherence} (uniform superposition of all states with exactly $N_p$ qubits in state $\ket{1}$ out of $N_T+N_p$ qubits). The Dicke state naturally encodes all configurations in which the $N_p$ projectile particles can be mapped onto the $N_T + N_p$ available positions, thereby providing a compact way to select and parallelize the required one-particle exchange operations without explicitly constructing the full set of many-particle permutations. Thus, when combined with controlled-swap operations between the target and projectile registers, the Dicke-state ancilla efficiently selects the appropriate exchange channels without the need to construct a separate circuit for each permutation, and implements explicitly only one-particle exchanges. The sign associated with each permutation can be straightforwardly incorporated by applying single-qubit $Z$ gates to the first $N_T$ ancilla qubits, which are entangled with an exchange of a target state with a projectile state. The subsequent uncomputation of the ancilla register is also simple, as it will only consist of $N_T+N_p$ \textsc{cnot}s. Several efficient quantum circuits for generating Dicke states exist in the literature~\cite{BaertschiEidenbenz2019Dicke,Mukherjee2020ActualDicke,BaertschiEidenbenz2022ShortDepth,YuanZhang2025DepthEfficientDicke},  making this approach practically feasible for fault-tolerant quantum hardware.

\begin{figure*}
    \centering
    \includegraphics[width=0.98\linewidth,clip]{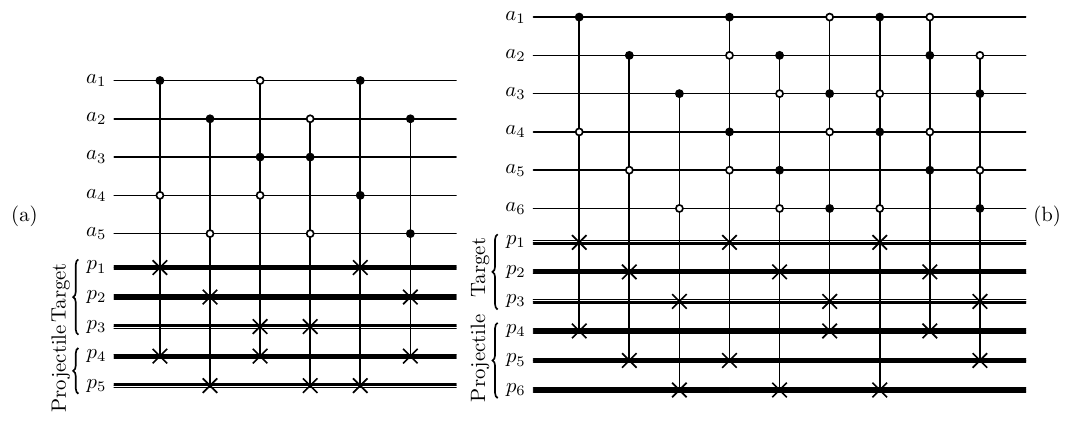}
    \caption{Single-particle swap operations generated by the antisymmetrization algorithm for (a) a system with two projectile particles and three target particles, and (b) a system with three particles in both the target and the projectile. Additional ancilla qubits, not shown here, may be introduced to simplify the multi-controlled swaps or to allow more swaps to be executed in parallel.
}
    \label{fig:asym-3p2p}
\end{figure*}

Figure~\ref{fig:antisymm2} shows the quantum circuit for antisymmetrizing the target and the projectile, each consisting of two indistinguishable fermions. Assuming that one begins with the antisymmetrized two-body states for the target and the projectile (each possibly a superposition of Slater determinants) and 
with four ancilla qubits prepared in the Dicke state $D^4_2$, the operations in the blue-filled region perform one-particle state swaps whenever any of the first two ancillae are in the state $\ket{1}$ and any of the last two are in the state $\ket{0}$. These operations generate three four-particle states in which at least one projectile single-particle state has been exchanged with a target single-particle state. The first two swaps generate all required exchanges between target and projectile 
states of the same particle rank. The final two swaps ensure exchanges between target and projectile particle 
states of different rank (e.g., the state of particle 1 in the target is exchanged 
with that of particle 2 in the projectile). Implementing these swaps individually as multi-controlled operations requires substantially more resources than an alternative construction in which additional ancillae (not shown) are first flipped based on the states of $a_1$ through $a_4$, as in the controlled operations of Fig.~\ref{fig:antisymm2}. In that case, each swap can be controlled by a single extra ancilla, which can then be uncomputed by the same multi-controlled operation, see Fig. 1 in Ref.~\cite{som}. If as many as $N_p$ additional ancillae are available, $N_p$ swap operations can be carried out in parallel; even if fewer qubits are available, the circuit depth can still be reduced relative to the fully multi-controlled implementation. The $Z$ gates applied only to the first two ancilla qubits provide the correct antisymmetrization phase, since any odd number of particle exchanges produces a $-1$ phase, while an even number of exchanges yields an overall phase of $+1$, as required. After all swaps are completed, any ancilla $a_i$ in state $\ket{1}$ is always entangled with the projectile single-particle state occupying the $i$th position. Thus, after the blue-filled region, we have generated all relevant permutations between the particles, with projectile states correlated with ancillae in state $\ket{1}$ and target states correlated with ancillae in state $\ket{0}$. The uncomputation procedure is shown in the pink-filled region of Fig.~\ref{fig:antisymm2}; as noted earlier, it consists of $N_T+N_p$ \textsc{cnot} operations. For the definition of these \textsc{cnot} gates, see Fig. \ref{fig:equations-cnot}(b). I note that although exponentially suppressed, the wave-packet tails might cross the midpoint of the box, leaving the ancilla register entangled with the system after uncomputation and introducing an error in the subsequent evolution. Although extremely unlikely, such errors can be detected by measuring the ancillae: any outcome containing a $\ket{1}$ indicates unsuccessful antisymmetrization and requires the state preparation to be repeated.

Each swap exchanges entire one-particle registers and therefore does not alter the internal antisymmetry of the initial target and projectile states. The circuit generates all distinct assignments of particles between the two subsystems with the corresponding permutation signs. An exchange of two particles originating in the same subsystem produces a minus sign from the initial antisymmetry, whereas an exchange between subsystems maps one term of the superposition onto another with the opposite sign. The resulting state therefore changes sign under any particle exchange and is fully antisymmetric. A numerical validation is provided in Ref.~\cite{github-implementation}.

A standard swap operation decomposes into three \textsc{cnot} gates, whereas a 
multi-controlled swap requires only two \textsc{cnot} gates together with a single multi-controlled \textsc{not} operation. Recent progress has shown that multi-controlled \textsc{not} can be very efficiently implemented on fault-tolerant hardware \cite{multiToff}. In turn, the modest cost of the controlled-swap operations used here advances the feasibility of simulating nuclear reactions on quantum hardware. Nevertheless, the full gate cost cannot be estimated in general because the required control operations depend on the specific projectile-target system.

In Fig. \ref{fig:asym-3p2p}, I show all the single-particle swaps necessary to antisymmetrize the target and projectile in a couple of situations: (a) two fermions in the projectile, and three in the target, and (b) three identical fermions in both the target and the projectile. The construction of the Dicke state, the phase application, which requires $Z$-gate operations on the first $N_T$ ancillae, and the uncomputing of the ancillae are not shown, but their execution is relatively simple. A general algorithm for antisymmetrizing $N_T$ particles in the target and $N_p$ particles in the projectile requires $N_T+N_p$ ancillae prepared in  Dicke state $D^{N_T+N_p}_{\,N_p}$, and then $N_TN_p$ multi-controlled one-state swaps between the target and projectile. This scaling follows from the need to account for all possible exchanges between the target and projectile states.  Note that all swaps should be set up in such a way that they do not exchange any of the states in both the target and the projectile that have already been previously exchanged. This can be accomplished by introducing additional controls. For example, the third swap in Fig.~\ref{fig:asym-3p2p}(a) includes an additional control on ancilla $a_1$ being in state $\ket{1}$. Without this control, if the swap were conditioned only on $a_3$ being in state $\ket{1}$ and $a_5$ being in the 
$\ket{0}$ state, some components of the wave function would simply undergo an exchange of the states in $p_1$ and $p_3$. Since these states are already antisymmetrized with respect to one another, such an exchange does not generate a new contribution to the antisymmetrized wave function.

The order of the swaps is irrelevant because each implements a distinct permutation and produces a different state. However, swaps can be executed in parallel only if they act on disjoint particle registers. As illustrated in Fig.~ \ref{fig:asym-3p2p}(a), circuits designed for smaller particle numbers can be reused to simplify the construction. In this case,
the first two and the last two swaps are following the swaps implemented for the 2+2 case presented in Fig. \ref{fig:antisymm2}, and this can be done similarly for other cases in which the target has two identical fermions in the projectile (see Figs. 2 and 3 in \cite{som}). The middle two swaps antisymmetrize the third particle in the target against the two particles in the projectile. 
%The correct antisymmetrization phase is incorporated through a small number of single-qubit $Z$ operations on the ancilla register, after which the ancilla qubits are uncomputed through controlled-\textsc{not} gates conditioned on the particle labels. 

This paper presents an efficient algorithm for constructing fully antisymmetric fermionic states from the product of two independently antisymmetrized subsystems, such as a target nucleus and a projectile, as required in nuclear reaction simulations. The method, designed as a complementary rather than competing approach to existing algorithms, employs a Dicke-state ancilla register to compactly encode all relevant one-particle exchange channels between the two subsystems, ensuring that only single-particle swaps are performed. The ancilla qubits select which exchanges are activated, allowing the required swaps to be executed in parallel. From several examples presented in this paper and the supplement \cite{som}, the algorithm requires $N_T N_p$ single-particle swaps (some of them collapsed using a quantum implementation of the \textsc{or} operation), motivated by the fact that all projectile states need to be swapped against $N_T$ particle states in the target. By introducing $N_p$ additional clean ancillae, up to $N_p$ swaps can be carried out simultaneously, significantly reducing the circuit depth relative to a purely sequential implementation.

%The correct antisymmetrization phase is incorporated through a small number of single-qubit $Z$ operations on the ancilla register, after which the ancilla qubits are efficiently disentangled. 
This approach avoids the need to apply inverse state preparation circuits for the target single-particle states, further simplifying the algorithm. However, we can devise a version of the antisymmetrization algorithm that uses a similar uncomputing algorithm as in Ref.~\cite{Rule:2025asym}, involving the inverse of the circuit that constructs the single particle states. In this case, one would need to implement $N_p$ inverse circuits instead of one for the uncomputation of each ancilla for each projectile and target state. Since the disentanglement of the ancillae is the most expensive part of the algorithm \cite{Rule:2025asym}, the benefit of performing parallel exchanges will quickly be offset by the application of state preparation circuits (both direct and inverse). Nevertheless, the present approach is well suited for antisymmetrizing proton and 
neutron wave functions, in an isospin formalism. Even when the subsystems are not strictly well separated, but some of the single-particle states can be efficiently distinguished, the parallel structure of the algorithm can provide a significant reduction in circuit depth when combined with the recursive antisymmetrization scheme of Ref.~\cite{Rule:2025asym} or any of the other existing algorithms~\cite{Berry2018a,Babbush:2023sas,10.1063/5.0239980}.

The parallelized structure of the algorithm makes it well-suited for implementation on fault-tolerant quantum hardware, enabling scalable quantum simulations of nuclear and chemical  reactions using a simple lattice representation, as proposed in Ref.~\cite{Rule:2025scatt}. In quantum chemistry, the same approach can be used to antisymmetrize electrons belonging to two spatially separated systems with respect to one another. For nuclear physics, the quantum circuits provided in this article and in the accompanying supplement \cite{som} allow target-projectile antisymmetrization for simulation of neutron-, proton-, deuteron-, tritium-, $^{3,4}$He-induced reactions for targets up to $^{12}$C.

\bigskip

\textit{Acknowledgments} 
I thank J. Carlson and M. Angel for stimulating discussions regarding the antisymmetrization of fermionic systems, E. Rule and C. Johnson for feedback on the manuscript, and A. Roggero for pointing out Ref.~\cite{multiToff}. This work was carried out under the auspices of the National Nuclear Security Administration of the U.S. Department of Energy at Los Alamos National Laboratory under Contract No. 89233218CNA000001.  The author gratefully acknowledges support by the Advanced Simulation and Computing (ASC) program; ChatGPT (OpenAI) was used for assistance with language 
editing and improving the clarity of the manuscript.

\bibliography{references,Dicke-refs}

\end{document}

\title{Antisymmetrization of composite fermionic states for quantum simulations of nuclear reactions in first-quantization mapping 
(supplemental online material)}

  \author{Ionel Stetcu}
\affiliation{
  Theoretical Division, Los Alamos National Laboratory, Los Alamos, New Mexico 87545, USA}

  \begin{abstract}
      Additional online material showing the parallelization of non-overlapping swaps, and the implementation of the antisymmetrization quantum circuit for other examples involving two identical fermions in the target.
  \end{abstract}
%\date{\today}
%\preprint{LA-UR-25-XXXXX}

\maketitle

\begin{figure}
    \centering
    \includegraphics[width=0.8\linewidth]{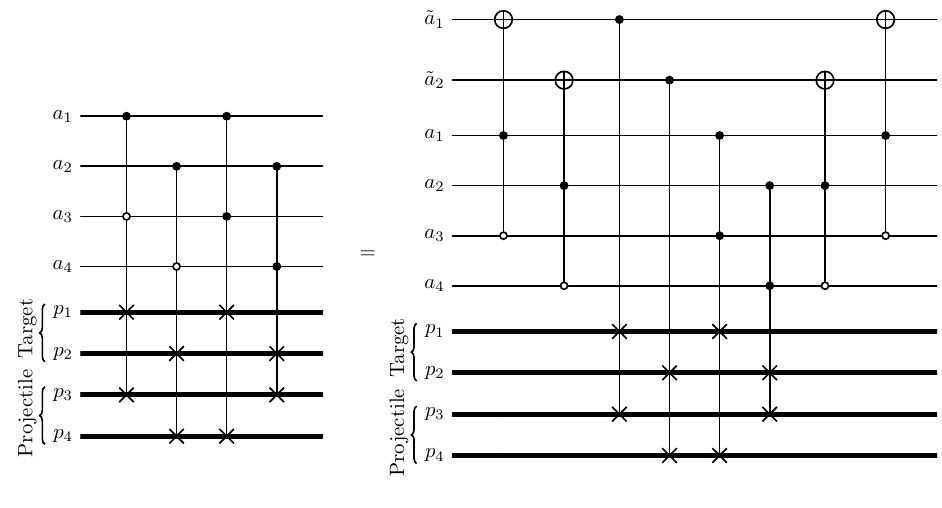}
    \caption{Parallelization of swap operations in Fig. 2 of the main paper with the help of two additional ancillae, which are uncomputed by the last two multi-controlled \textsc{not} gates. The same type of operations can be implemented to parallelize pairs of swaps in Fig. 3(a) in the main paper, and Figs. \ref{fig:qcirc-4p2p} and \ref{fig:qcirc-6p2p}, as well as triplets of swaps in Fig. 3(b) using three additional ancillae.}
    \label{fig:qcirc-2p2p-p}
\end{figure}

\begin{figure}
    \centering
    \includegraphics[width=0.7\linewidth]{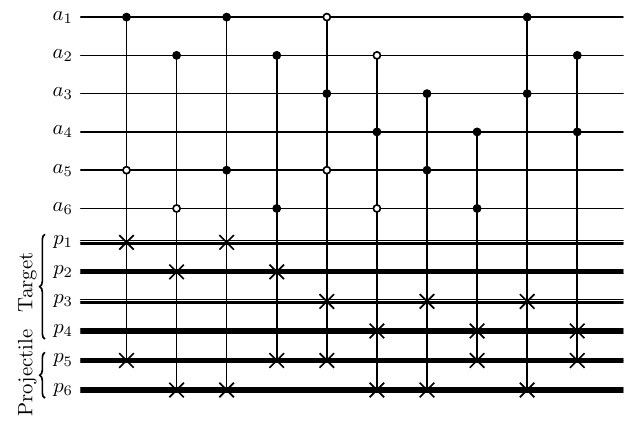}
    \caption{Single-particle swap operations generated by the antisymmetrization algorithm for a system with two projectile particles and four target particles. Swaps 7 and 9 can be implemented as a single swap, using Eq. (\ref{eq:swap_collapse}) which requires an additional ancilla (and similarly for 8 and 10, either by reusing the ancilla for 7 and 10, or in parallel with the aid of another ancilla). In total, 8 independent swaps need to be performed, and four additional ancillae allow the parallelization. As noted in the main paper, $N_p=2$ parallel swaps can be implemented using at most four ancillae.}
    \label{fig:qcirc-4p2p}
\end{figure}

\begin{figure}
    \centering
    \includegraphics[width=0.8\linewidth]{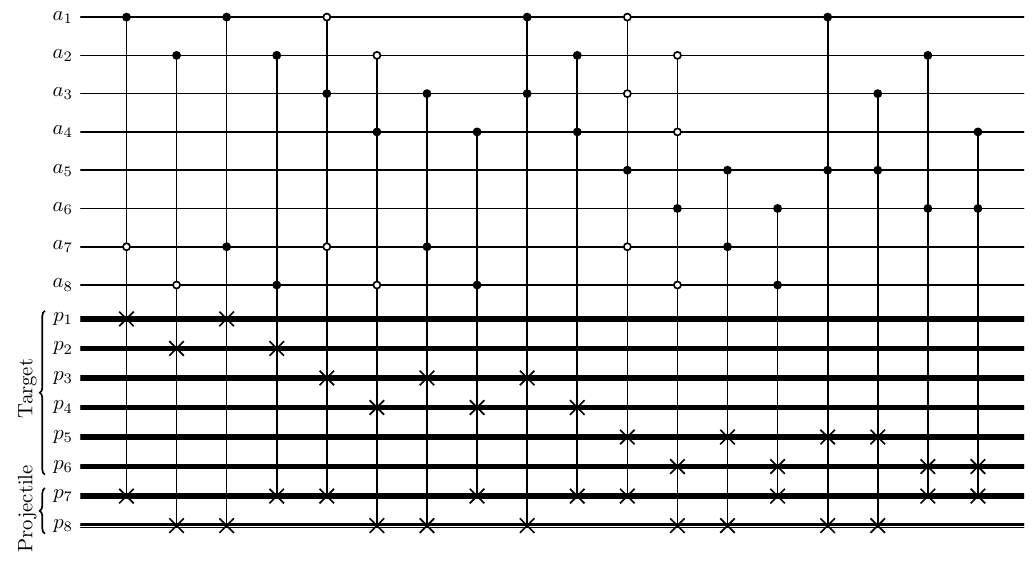}
    \caption{Same as in Fig.~\ref{fig:qcirc-4p2p}, but for six identical fermions in the target. Swaps 7 and 9 can be collapsed again using Eq.~(\ref{eq:swap_collapse}), and similarly for 8 and 10. Implementing a similar circuit to Eq. (\ref{eq:swap_collapse}), one can collapse swaps 13, 15 and 16, and similarly for swaps 14, 17 and 18. Hence, as noted in the main paper's abstract, 12 independent swaps can be implemented in parallel.}
    \label{fig:qcirc-6p2p}
\end{figure}

The following circuit implements the equivalent of \textsc{or} in order to implement two swaps in parallel:

\begin{equation}
\begin{quantikz}[row sep=0.7cm, column sep=0.7cm]
%------------------ Ancillae ------------------
 \lstick{$a_1$}  &
\ctrl{2}&&\\
\lstick{$a_2$}  &  
&\ctrl{1}&\\
\lstick{$a_3$} & 
 \ctrl{1}&\ctrl{1}&\\
%------------------ Particle 1 (2 qubits) ------------------
\lstick{$p_1$} \setwiretype{b}   & 
\swap{1}&\swap{1}&\\
%------------------ Particle 2 (2 qubits) ------------------
\lstick{$p_2$}\setwiretype{b} & 
\targX{}&\targX{}&
\end{quantikz} \quad
=
\quad
\begin{quantikz}[row sep=0.7cm, column sep=0.7cm]
%------------------ Ancillae ------------------
\lstick{$a_0$} & 
\targ{}& \octrl{3}&\targ{}&\\
 \lstick{$a_1$}  &
\octrl{-1}&&\octrl{-1}&\\
\lstick{$a_2$}  &  
\octrl{-1}&&\octrl{-1}&\\
\lstick{$a_3$} & 
&\ctrl{1}&&\\
%------------------ Particle 1 (2 qubits) ------------------
\lstick{$p_1$} \setwiretype{b}   & 
&\swap{1}&&\\
%------------------ Particle 2 (2 qubits) ------------------
\lstick{$p_2$}\setwiretype{b} & 
&\targX{}&&
\end{quantikz}
\label{eq:swap_collapse}
\end{equation}.